\documentclass[3p,times,twocolumn]{elsarticle}

\usepackage{ecrc}

\volume{00}

\firstpage{1}

\journalname{Nuclear Physics B Proceedings Supplements}

\runauth{O. Yasuda}

\jnltitlelogo{Nuclear Physics B Proceedings Supplements}

\usepackage{amssymb}
\usepackage[figuresright]{rotating}

\begin{document}

\begin{frontmatter}


%\title{Some constraints on new physics by atmospheric neutrinos\tnoteref{label1}}
%\author{Osamu Yasuda\corref{cor1}\fnref{label2}}
%\address{Department of Physics, Tokyo Metropolitan University,
%Minami-Osawa, Hachioji, Tokyo 192-0397, Japan\fnref{label3}}

\dochead{}

\title{Some constraints on new physics by atmospheric neutrinos}


\author{Osamu Yasuda}

\address{Department of Physics, Tokyo Metropolitan University,
Minami-Osawa, Hachioji, Tokyo 192-0397, Japan}

\begin{abstract}

From the analytic oscillation
probability for high energy atmospheric neutrinos
in the presence of new physics in propagation,
it is argued that the components
$|\epsilon_{e\mu}|$ and $|\epsilon_{\mu\tau}|$ as well as
the quantity
$|\epsilon_{\tau\tau}-|\epsilon_{e\tau}|^2/(1+\epsilon_{ee})|$
should be small.
\end{abstract}


\end{frontmatter}

% \linenumbers


There has been much interest in the
effective Non-Standard neutral current-neutrino Interactions (NSI)
with matter, because the future long baseline experiments are expected
to allow us to probe them.  With such interactions,
the matter potential in the basis of the flavor eigenstates
becomes a $3 \times 3$ hermitian matrix
\begin{eqnarray}
{\cal A}_{\alpha\beta}=
%\left(\begin{array}{ccc}
A(\delta_{e\alpha}\delta_{e\beta}
+ \epsilon_{\alpha\beta}),\quad(\alpha,\beta=e,\mu,\tau)
\label{matter-np}
\end{eqnarray}
where $A\equiv\sqrt{2}G_FN_e$ and $N_e$ is the density
of electrons in matter.
Constraints on the NSI parameters
$\epsilon_{\alpha\beta}$
from various neutrino experiments have been discussed
by many people (see Ref.\,\cite{Oki:2010uc} and references therein),
and the constraints at 90\%CL
are summarized by the following\,\cite{Biggio:2009nt}:
$|\epsilon_{ee}| \lesssim 4\times 10^0$,
$|\epsilon_{e\mu}| \lesssim 3\times 10^{-1}$,
$|\epsilon_{e\tau}| \lesssim 3\times 10^{0\ }$,
$|\epsilon_{\mu\mu}| \lesssim 7\times 10^{-2}$,
$|\epsilon_{\mu\tau}| \lesssim 3\times 10^{-1}$,
$|\epsilon_{\tau\tau}| \lesssim 2\times 10^{1\ }$.

Let us now consider the high-energy behavior
of the disappearance probability
$P(\nu_\mu\rightarrow\nu_\mu)$,
which can be measured by the upward-going
$\mu$ events in the atmospheric neutrino
experiments, and let us expand it in
$\Delta E/A\equiv\Delta m^2_{atm}/2AE$
($E$ is the neutrino energy,
$\Delta m^2_{atm}$ is the mass
squared-difference of the atmospheric
neutrino oscillation):
\begin{eqnarray}
1-P(\nu_\mu\rightarrow\nu_\mu)\simeq
c_0 + c_1\frac{\Delta E}{A}
+ {\cal O}\left(\left(\frac{\Delta E}{A}\right)^2\right).
\label{expansion}
\end{eqnarray}
It is known that
$c_0$ and $c_1$ in Eq.\,(\ref{expansion})
vanish in the standard scenario
with three flavors, and the atmospheric neutrino data
of Superkamiokande indeed confirms
that $|c_0|\ll 1$ and $|c_1|\ll 1$.

In the presence of the generic matter potential
(\ref{matter-np}), it can be shown \cite{Oki:2010uc} that
$|c_0|\ll 1$ leads to 
$|\epsilon_{e\mu}|\ll1$,
$|\epsilon_{\mu\mu}|\ll1$
and $|\epsilon_{\mu\tau}|\ll1$,
while $|c_1|\ll 1$ leads to
$|\epsilon_{\tau\tau}-|\epsilon_{e\tau}|^2/(1+\epsilon_{ee})|\ll1$.
$|\epsilon_{\mu\mu}|\ll1$, $|\epsilon_{\mu\tau}|\ll1$ and
$|\epsilon_{\tau\tau}-|\epsilon_{e\tau}|^2/(1+\epsilon_{ee})|\ll1$ were
first shown in Refs.\,\cite{Davidson:2003ha}, \cite{Fornengo:2001pm}
and \cite{Friedland:2004ah},
respectively.  On the other hand,
if we exclude the one-loop
arguments\,\cite{Davidson:2003ha} to constrain $\epsilon_{e\mu}$
as in Ref.\,\cite{Biggio:2009kv}, then
the observation in Ref.\,\cite{Oki:2010uc} that $|\epsilon_{e\mu}|\ll1$ follows
from the the atmospheric neutrino constraint is new,
although it is based only on an analytical treatment.

It would be interesting if we can
determine the coefficient of the term of order
$(\Delta E/A)^2$ in Eq.\,(\ref{expansion}) and its zenith
angle dependence
by experiments such as Superkamiokande,
IceCube and Hyperkamiokande.

This research was partly supported by a Grant-in-Aid for Scientific
Research of the Ministry of Education, Science and Culture, under
Grant No. 21540274.



%% References
\vglue 0.5cm

\end{document}